\documentclass[ aip,  jmp,  preprint]{revtex4-1}
\usepackage{amsmath}
\usepackage{graphicx}
\usepackage{dcolumn}
\usepackage{bm}

\begin{document}

\preprint{}
\title[]{Equation satisfied by electron-electron mutual Coulomb repulsion energy
density functional.}
\author{Daniel P. Joubert}
\email{daniel.joubert2@wits.ac.za}
\affiliation{Centre for Theoretical Physics, University of the Witwatersrand, PO Wits
2050, Johannesburg, South Africa}
\date{\today }

\begin{abstract}
It is shown that the electron-electron mutual Coulomb repulsion energy
density functional $V_{ee}^{\gamma }\left[ \rho \right] $ satisfies the
equation
\begin{eqnarray*}
&&V_{ee}^{\gamma }\left[ \rho _{N}^{1}\right] -V_{ee}^{\gamma }\left[ \rho
_{N-1}^{\gamma }\right] \\
&=&\int d^{3}r\frac{\delta V_{ee}^{\gamma }\left[ \rho _{N}^{1}\right] }{%
\delta \rho _{N}^{1}\left( \mathbf{r}\right) }\left( \rho _{N}^{1}\left(
\mathbf{r}\right) -\rho _{N-1}^{\gamma }\left( \mathbf{r}\right) \right)
\end{eqnarray*}%
where $\rho _{N}^{1}\left( \mathbf{r}\right) $ and $\rho _{N-1}^{\gamma
}\left( \mathbf{r}\right) $ are $N$-electron and $(N-1)$-electron densities
determined from the same adiabatic scaled external potential of the $N$%
-electron system at coupling strength $\gamma .$
\end{abstract}

\maketitle









\section{ Introduction}

Density Functional Theory (DFT) \cite{HohenbergKohn:64} is one of the most
important tools for the calculation of electronic structure and structural
properties of solids and molecules\cite{AdriennRuzsinszky2011}. In all
practical applications approximations are used\cite{AdriennRuzsinszky2011}.
It is therefore important to examine the exact formal properties of
functionals that enter into the formalism as an aid to understanding the
underlying formulation and to develop improved approximations. Here the
Coulomb interaction energy functional%
\begin{equation}
V_{ee}^{\gamma }\left[ \rho _{N}^{\gamma }\right] =\left\langle \Psi _{\rho
_{N}^{\gamma }}^{\gamma }\left\vert \hat{V}_{ee}\right\vert \Psi _{\rho
_{N}^{\gamma }}^{\gamma }\right\rangle
\end{equation}%
is examined. $\left\vert \Psi _{\rho _{N}^{\gamma }}^{\gamma }\right\rangle $
is the $N$-electron wavefunction that yields the density $\rho _{N}^{\gamma
} $ and minimizes $\left\langle \Psi \left\vert \hat{T}+\gamma \hat{V}%
_{ee}\right\vert \Psi \right\rangle $\cite{Levy:79}$:$
\begin{equation}
\min_{\Psi \rightarrow \rho _{N}^{\gamma }}\left\langle \Psi \left\vert \hat{%
T}+\gamma \hat{V}_{ee}\right\vert \Psi \right\rangle =\left\langle \Psi
_{\rho _{N}^{\gamma }}^{\gamma }\left\vert \hat{V}_{ee}\right\vert \Psi
_{\rho _{N}^{\gamma }}^{\gamma }\right\rangle .  \label{i2}
\end{equation}%
$\hat{T}$ and $\hat{V}_{ee}$ are the kinetic energy and electron-electron
repulsion operators respectively and $\gamma $ scales the electron-electron
interaction energy. In this paper it is shown that

\begin{eqnarray}
&&V_{ee}^{\gamma }\left[ \rho _{N}^{1}\right] -V_{ee}^{\gamma }\left[ \rho
_{N-1}^{\gamma }\right]  \notag \\
&=&\int d^{3}r\frac{\delta V_{ee}^{\gamma }\left[ \rho _{N}^{1}\right] }{%
\delta \rho _{N}^{1}\left( \mathbf{r}\right) }\left( \rho _{N}^{1}\left(
\mathbf{r}\right) -\rho _{N-1}^{\gamma }\left( \mathbf{r}\right) \right) .
\label{vee}
\end{eqnarray}%
where $\rho _{N}^{1}\left( \mathbf{r}\right) $ and $\rho _{N-1}^{\gamma
}\left( \mathbf{r}\right) $ are $N$-electron and $(N-1)$-electron densities
determined from the same external adiabatic potential at coupling strength $%
\gamma $ as discussed below. As a corollary it will be shown that%
\begin{equation}
V_{ee}^{\gamma }\left[ \rho _{N}^{1}\right] =\sum_{L=0}^{N-2}\int d^{3}r%
\frac{\delta V_{ee}^{\gamma }\left[ \rho _{N-L}^{\gamma }\right] }{\delta
\rho _{N-L}^{\gamma }\left( \mathbf{r}\right) }\left( \rho _{N-L}^{\gamma
}\left( \mathbf{r}\right) -\rho _{N-1-L}^{\gamma }\left( \mathbf{r}\right)
\right) .  \label{vee_c}
\end{equation}

\section{Proof}

In the adiabatic connection approach \cite%
{HarrisJones:74,LangrethPerdew:75,LangrethPerdew:77,GunnarsonLundqvist:76}
of the constrained minimization formulation of density functional theory
\cite{HohenbergKohn:64,KohnSham:65,Levy:79,LevyPerdew:85} the Hamiltonian $%
\hat{H}^{\gamma }$ for a system of $N$ electrons is given by
\begin{equation}
\hat{H}_{N}^{\gamma }=\hat{T}^{N}+\gamma \hat{V}_{ee}^{N}+\hat{v}_{N,\text{%
ext}}^{\gamma }\left[ \rho _{N}\right] .  \label{a3}
\end{equation}%
Atomic units, $\hbar =e=m=1$ are used throughout. $\hat{T}$ is the kinetic
energy operator,%
\begin{equation}
\hat{T}^{N}=-\frac{1}{2}\sum_{i=1}^{N}\nabla _{i}^{2},  \label{a4}
\end{equation}%
\ and $\gamma \hat{V}_{\text{ee }}$is a scaled electron-electron interaction,%
\begin{equation}
\gamma \hat{V}_{ee}^{N}=\gamma \sum_{i<j}^{N}\frac{1}{\left\vert \mathbf{r}%
_{i}-\mathbf{r}_{j}\right\vert }.  \label{a2}
\end{equation}%
The the external potential
\begin{equation}
\hat{v}_{N,\text{ext}}^{\gamma }\left[ \rho _{N}\right] =\sum_{i=1}^{N}v_{%
\text{ext}}^{\gamma }\left( \left[ \rho _{N}\right] ;\mathbf{r}_{i}\right) ,
\label{a1}
\end{equation}%
is constructed to keep the charge density fixed at $\rho _{N}\left( \mathbf{r%
}\right) ,$ the ground state charge density of the fully interacting system (%
$\gamma =1$), for all values of the coupling constant $\gamma .$ The
external potential has the form \cite{LevyPerdew:85,GorlingLevy:93}
\begin{align}
v_{\text{ext}}^{\gamma }(\left[ \rho _{N}\right] ;\mathbf{r})& =\left(
1-\gamma \right) v_{ux}([\rho _{N}];\mathbf{r})  \notag \\
& +v_{c}^{1}([\rho _{N}];\mathbf{r)}-v_{c}^{\gamma }([\rho _{N}];\mathbf{r)+}%
v_{\text{ext}}^{1}(\left[ \rho _{N}\right] ;\mathbf{r}),  \label{e1}
\end{align}%
where $v_{\text{ext}}^{1}(\left[ \rho _{N}\right] ;\mathbf{r})=v_{\text{ext}%
}\left( \mathbf{r}\right) $ is the external potential at full coupling
strength, $\gamma =1,$ and $v_{\text{ext}}^{0}(\left[ \rho _{N}\right] ;%
\mathbf{r})$ is non-interacting Kohn-Sham potential. The exchange plus
Hartree potential \cite{ParrYang:bk89,DreizlerGross:bk90}$v_{ux}([\rho _{N}];%
\mathbf{r}),$ is independent of $\gamma ,$ while the correlation potential $%
v_{c}^{\gamma }([\rho _{N}];\mathbf{r)}$ depends in the scaling parameter $%
\gamma .$

\bigskip The chemical potential%
\begin{equation}
\mu _{N}=E_{N}^{\gamma }\left( v_{\text{ext}}^{\gamma }\left[ \rho _{N}%
\right] \right) -E_{N-1}^{\gamma }\left( v_{\text{ext}}^{\gamma }\left[ \rho
_{N}\right] \right)  \label{e2}
\end{equation}%
depends on the asymptotic decay of the charge density \cite%
{ParrYang:bk89,DreizlerGross:bk90,JonesGunnarsson:89}, and hence is
independent of the coupling constant $\gamma $ \cite%
{LevyGorling:96,LevyGorlingb:96}. In Eq. (\ref{e2}) $E_{N-1}^{\gamma }$ is
the groundstate energy of the $\left( N-1\right) $-electron system with the
same single-particle external potential $v_{\text{ext}}^{\gamma }\left( %
\left[ \rho _{N}\right] ;\mathbf{r}\right) $ as the $N$-electron system:%
\begin{eqnarray}
\hat{H}_{N-1}^{\gamma }\left\vert \Psi _{\rho _{N-1}^{\gamma }}^{\gamma
}\right\rangle &=&E_{N-1}^{\gamma }\left\vert \Psi _{\rho _{N-1}^{\gamma
}}^{\gamma }\right\rangle  \notag \\
\hat{H}_{N-1}^{\gamma } &=&\hat{T}^{N-1}+\gamma \hat{V}_{\text{ee}}^{N-1}+%
\hat{v}_{N-1,\text{ext}}^{\gamma }\left[ \rho _{N}\right]  \notag \\
\hat{v}_{N-1,\text{ext}}^{\gamma }\left[ \rho _{N}\right] &=&%
\sum_{i=1}^{N-1}v_{\text{ext}}^{\gamma }\left( \left[ \rho _{N}\right] ;%
\mathbf{r}_{i}\right)  \label{h1}
\end{eqnarray}%
Note that by construction of $v_{\text{ext}}^{\gamma }\left( \left[ \rho _{N}%
\right] ;\mathbf{r}\right) ,$ Eq. (\ref{e1}) $\rho _{N}=\rho _{N}^{1}$ is
independent of $\gamma ,$ but the groundstate density of the $\left(
N-1\right) $-electron system $\rho _{N-1}^{\gamma }$, is a function of $%
\gamma .$

The correlation energy $E_{c}^{\gamma }\left[ \rho _{N-1}^{\gamma }\right] $
is defined as\cite{LevyPerdew:85}
\begin{eqnarray}
E_{c}^{\gamma }\left[ \rho _{N-1}^{\gamma }\right] &=&\left\langle \Psi
_{\rho _{N-1}^{\gamma }}^{\gamma }\left\vert \hat{T}^{N-1}+\gamma \hat{V}%
_{ee}^{N-1}\right\vert \Psi _{\rho _{N-1}^{\gamma }}^{\gamma }\right\rangle
\notag \\
&&-\left\langle \Psi _{\rho _{N-1}^{\gamma }}^{0}\left\vert \hat{T}%
^{N-1}+\gamma \hat{V}_{ee}^{N-1}\right\vert \Psi _{\rho _{N-1}^{\gamma
}}^{0}\right\rangle ,  \label{ec1}
\end{eqnarray}%
where $\left\vert \Psi _{\rho _{N-1}^{\gamma }}^{0}\right\rangle $ is the
Kohn-Sham $\left( N-1\right) $ independent particle groundstate wavefunction
that yields the same density as the interacting $\left( N-1\right) $%
-electron system at coupling strength $\gamma .$ The correlation part of the
kinetic energy is given by%
\begin{eqnarray}
T_{c}^{\gamma }\left[ \rho _{N-1}^{\gamma }\right] &=&\left\langle \Psi
_{\rho _{N-1}^{\gamma }}^{\gamma }\left\vert \hat{T}^{N-1}\right\vert \Psi
_{\rho _{N-1}^{\gamma }}^{\gamma }\right\rangle  \notag \\
&&-\left\langle \Psi _{\rho _{N-1}^{\gamma }}^{0}\left\vert \hat{T}%
^{N-1}\right\vert \Psi _{\rho _{N-1}^{\gamma }}^{0}\right\rangle .
\label{ec2}
\end{eqnarray}

The virial theorem can be written as\cite{LevyPerdew:85}

\begin{equation}
2T^{\gamma }\left[ \rho \right] +\gamma V_{ee}+\int d^{3}r^{\prime
}v^{\gamma }\left( \left[ \rho \right] ;\mathbf{r}^{\prime }\right) \left(
3\rho \left( \mathbf{r}^{\prime }\right) +\mathbf{r}^{\prime }\mathbf{%
.\nabla }^{\prime }\rho \left( \mathbf{r}^{\prime }\right) \right) =0
\label{v1}
\end{equation}%
or as
\begin{equation}
T^{\gamma }\left[ \rho \right] +F^{\gamma }\left[ \rho \right] +\int
d^{3}r^{\prime }v^{\gamma }\left( \left[ \rho \right] ;\mathbf{r}^{\prime
}\right) \left( 3\rho \left( \mathbf{r}^{\prime }\right) +\mathbf{r}^{\prime
}\mathbf{.\nabla }^{\prime }\rho \left( \mathbf{r}^{\prime }\right) \right)
=0  \label{v2}
\end{equation}%
where the energy functional $F^{\gamma }\left[ \rho \right] $ is given by
\cite{ParrYang:bk89,DreizlerGross:bk90}%
\begin{eqnarray}
F^{\gamma }\left[ \rho \right] &=&T^{\gamma }\left[ \rho \right] +\gamma
V_{ee}^{\gamma }\left[ \rho \right]  \notag \\
&=&T^{0}\left[ \rho \right] +\gamma \left( E_{x}\left[ \rho \right] +U\left[
\rho \right] \right) +E_{c}^{\gamma }\left[ \rho \right]  \label{v3}
\end{eqnarray}
$T^{\gamma }\left[ \rho \right] $ and $E_{c}^{\gamma }\left[ \rho \right] $
are the kinetic and correlation energies for the interacting system at
coupling strength $\gamma $ and density $\rho $ while $E_{x}\left[ \rho %
\right] $ and $U\left[ \rho \right] $ are the exchange and Hartree energies
at $\rho .$

Take the functional derivative of Eq. (\ref{v2}) with respect to $v^{\gamma
}\left( \mathbf{r}^{\prime \prime }\right) =v_{\text{ext}}^{\gamma }\left( %
\left[ \rho \right] ;\mathbf{r}^{\prime \prime }\right) $ and use the chain
rule for functional derivatives (here and in the following the assumption is
made that all the functional derivatives are well defined \cite{PPLB:82}):
\begin{eqnarray}
0 &=&\int d^{3}r^{\prime }\frac{\delta T^{\gamma }\left[ \rho \right] }{%
\delta \rho \left( \mathbf{r}^{\prime }\right) }\left. \frac{\delta \rho
\left( \mathbf{r}^{\prime }\right) }{\delta v^{\gamma }\left( \mathbf{r}%
^{\prime \prime }\right) }\right\vert _{N}+\int d^{3}r^{\prime }\frac{\delta
F^{\gamma }\left[ \rho \right] }{\delta \rho \left( \mathbf{r}^{\prime
}\right) }\left. \frac{\delta \rho \left( \mathbf{r}^{\prime }\right) }{%
\delta v^{\gamma }\left( \mathbf{r}^{\prime \prime }\right) }\right\vert
_{N}+3\rho \left( \mathbf{r}^{\prime \prime }\right) +\mathbf{r}^{\prime
\prime }\mathbf{.\nabla }^{\prime \prime }\rho \left( \mathbf{r}^{\prime
\prime }\right)  \notag \\
&&-\int d^{3}r^{\prime }\mathbf{r}^{\prime }\mathbf{.\nabla }^{\prime
}v^{\gamma }\left( \left[ \rho \right] ;\mathbf{r}^{\prime }\right) \left.
\frac{\delta \rho \left( \mathbf{r}^{\prime }\right) }{\delta v^{\gamma
}\left( \mathbf{r}^{\prime \prime }\right) }\right\vert _{N}.  \label{v4}
\end{eqnarray}%
The subscript $N$ in $\left. \frac{\delta \rho \left( \mathbf{r}^{\prime
}\right) }{\delta v^{\gamma }\left( \mathbf{r}^{\prime \prime }\right) }%
\right\vert _{N}$ indicates that the functional derivative is taken at
constant particle number. A useful expression is the Berkowitz-Parr equation%
\cite{MaxBerkowitz1988,P.Fuentalba2007} which states that%
\begin{equation}
\left. \frac{\delta \rho \left( \mathbf{r}^{\prime }\right) }{\delta
v^{\gamma }\left( \mathbf{r}\right) }\right\vert _{N}=\left( -s^{\gamma
}\left( \mathbf{r}^{\prime },\mathbf{r}\right) +\frac{s^{\gamma }\left(
\mathbf{r}^{\prime }\right) s^{\gamma }\left( \mathbf{r}\right) }{S^{\gamma }%
}\right) ,  \label{bp}
\end{equation}%
where%
\begin{equation}
s^{\gamma }\left( \mathbf{r}^{\prime },\mathbf{r}\right) =\left( \frac{%
\delta ^{2}F^{\gamma }\left[ \rho \right] }{\delta \rho \left( \mathbf{r}%
^{\prime }\right) \delta \rho \left( \mathbf{r}\right) }\right) ^{-1},
\label{bp1}
\end{equation}%
\begin{equation}
s^{\gamma }\left( \mathbf{r}\right) =\int d^{3}rs^{\gamma }\left( \mathbf{r},%
\mathbf{r}^{\prime }\right)  \label{bp2}
\end{equation}%
and
\begin{equation}
S^{\gamma }=\frac{1}{\eta ^{\gamma }}  \label{bp3}
\end{equation}%
where%
\begin{equation}
\eta ^{\gamma }=\left. \frac{\partial \mu }{\partial N}\right\vert
_{v^{\gamma }}  \label{bp4}
\end{equation}%
and $\mu $ is the chemical potential, independent of $\gamma $\cite%
{LevyGorling:96,LevyGorlingb:96}$.$\bigskip\ In addition
\begin{equation}
s^{\gamma }\left( \mathbf{r}\right) =S^{\gamma }f^{\gamma }\left( \mathbf{r}%
\right)  \label{bp5}
\end{equation}%
where the Fukui function \cite{MaxBerkowitz1988,P.Fuentalba2007}%
\begin{equation}
f^{\gamma }\left( \mathbf{r}\right) =\left. \frac{\delta \mu }{\delta \rho
\left( \mathbf{r}\right) }\right\vert _{N}=\left. \frac{\delta \rho \left(
\mathbf{r}\right) }{\delta N}\right\vert _{v^{\gamma }}  \label{bp6}
\end{equation}%
satisfy%
\begin{equation}
\int d^{3}r^{\prime }\frac{\delta ^{2}F^{\gamma }\left[ \rho \right] }{%
\delta \rho \left( \mathbf{r}\right) \delta \rho \left( \mathbf{r}^{\prime
}\right) }f^{\gamma }\left( \mathbf{r}^{\prime }\right) =\eta ^{\gamma },
\label{bp7}
\end{equation}%
where $\eta ^{\gamma }$ is a constant.

Multiply (\ref{v4}) by $\frac{\delta ^{2}F^{\gamma }\left[ \rho \right] }{%
\delta \rho \left( \mathbf{r}^{\prime \prime }\right) \delta \rho \left(
\mathbf{r}\right) }$ and integrate over $\mathbf{r}^{\prime \prime }.$ With (%
\ref{bp}) and (\ref{bp1}) the result is

\begin{eqnarray}
0 &=&-\frac{\delta T^{\gamma }\left[ \rho \right] }{\delta \rho \left(
\mathbf{r}\right) }-\frac{\delta F^{\gamma }\left[ \rho \right] }{\delta
\rho \left( \mathbf{r}\right) }+\mathbf{r.\nabla }v^{\gamma }\left( \left[
\rho \right] ;\mathbf{r}\right)  \notag \\
&&+\int d^{3}r^{\prime \prime }\int d^{3}r^{\prime }\frac{\delta T^{\gamma }%
\left[ \rho \right] }{\delta \rho \left( \mathbf{r}^{\prime }\right) }\frac{%
s^{\gamma }\left( \mathbf{r}^{\prime }\right) s^{\gamma }\left( \mathbf{r}%
^{\prime \prime }\right) }{S^{\gamma }}\frac{\delta ^{2}F^{\gamma }\left[
\rho \right] }{\delta \rho \left( \mathbf{r}^{\prime \prime }\right) \delta
\rho \left( \mathbf{r}\right) }  \notag \\
&&+\int d^{3}r^{\prime \prime }\int d^{3}r^{\prime }\frac{\delta F^{\gamma }%
\left[ \rho \right] }{\delta \rho \left( \mathbf{r}^{\prime }\right) }\frac{%
s^{\gamma }\left( \mathbf{r}^{\prime }\right) s^{\gamma }\left( \mathbf{r}%
^{\prime \prime }\right) }{S^{\gamma }}\frac{\delta ^{2}F^{\gamma }\left[
\rho \right] }{\delta \rho \left( \mathbf{r}^{\prime \prime }\right) \delta
\rho \left( \mathbf{r}\right) }  \notag \\
&&-\int d^{3}r^{\prime \prime \prime }\int d^{3}r^{\prime }\int
d^{3}r^{\prime }\mathbf{r}^{\prime }\mathbf{.\nabla }^{\prime }v^{\gamma
}\left( \left[ \rho \right] ;\mathbf{r}^{\prime }\right) \frac{s^{\gamma
}\left( \mathbf{r}^{\prime }\right) s^{\gamma }\left( \mathbf{r}^{\prime
\prime }\right) }{S^{\gamma }}\frac{\delta ^{2}F^{\gamma }\left[ \rho \right]
}{\delta \rho \left( \mathbf{r}^{\prime \prime }\right) \delta \rho \left(
\mathbf{r}\right) }  \notag \\
&&+\int d^{3}r^{\prime }\left( 3\rho \left( \mathbf{r}^{\prime \prime
}\right) +\mathbf{r}^{\prime \prime }\mathbf{.\nabla }^{\prime \prime }\rho
\left( \mathbf{r}^{\prime \prime }\right) \right) \frac{\delta ^{2}F^{\gamma
}\left[ \rho \right] }{\delta \rho \left( \mathbf{r}^{\prime \prime }\right)
\delta \rho \left( \mathbf{r}\right) }  \label{v5}
\end{eqnarray}%
Since $\frac{\delta ^{2}F^{\gamma }\left[ \rho \right] }{\delta \rho \left(
\mathbf{r}\right) \delta \rho \left( \mathbf{r}^{\prime }\right) }$ is
symmetric in $\mathbf{r}$ and $\mathbf{r}^{\prime },$ Eq. (\ref{v5}), with
the relations (\ref{bp5}) and (\ref{bp7}) becomes
\begin{eqnarray}
0 &=&-\frac{\delta T^{\gamma }\left[ \rho \right] }{\delta \rho \left(
\mathbf{r}\right) }-\frac{\delta F^{\gamma }\left[ \rho \right] }{\delta
\rho \left( \mathbf{r}\right) }+\mathbf{r.\nabla }v^{\gamma }\left( \left[
\rho \right] ;\mathbf{r}\right)  \notag \\
&&+\int d^{3}r^{\prime }\frac{\delta T^{\gamma }\left[ \rho \right] }{\delta
\rho \left( \mathbf{r}^{\prime }\right) }f^{\gamma }\left( \mathbf{r}%
^{\prime }\right)  \notag \\
&&+\int d^{3}r^{\prime }\frac{\delta F^{\gamma }\left[ \rho \right] }{\delta
\rho \left( \mathbf{r}^{\prime }\right) }f^{\gamma }\left( \mathbf{r}%
^{\prime }\right) -\int d^{3}r^{\prime }\mathbf{r}^{\prime }\mathbf{.\nabla }%
^{\prime \prime }v^{\gamma }\left( \left[ \rho \right] ;\mathbf{r}^{\prime
}\right) f^{\gamma }\left( \mathbf{r}^{\prime }\right)  \notag \\
&&+\int d^{3}r^{\prime }\left( 3\rho \left( \mathbf{r}^{\prime \prime
}\right) +\mathbf{r}^{\prime \prime }\mathbf{.\nabla }^{\prime \prime }\rho
\left( \mathbf{r}^{\prime \prime }\right) \right) \frac{\delta ^{2}F^{\gamma
}\left[ \rho \right] }{\delta \rho \left( \mathbf{r}^{\prime \prime }\right)
\delta \rho \left( \mathbf{r}\right) }.  \label{v6}
\end{eqnarray}

Since\cite{ParrYang:bk89}%
\begin{eqnarray}
&&\frac{d}{d\lambda }\frac{\delta G\left[ \rho _{\lambda }\right] }{\delta
\rho _{\lambda }\left( \mathbf{r}\right) }  \notag \\
&=&\int d^{3}r^{\prime \prime }\frac{\delta G\left[ \rho _{\lambda }\right]
}{\delta \rho _{\lambda }\left( \mathbf{r}\right) \delta \rho _{\lambda
}\left( \mathbf{r}^{\prime \prime }\right) }\frac{d}{d\lambda }\rho
_{\lambda }\left( \mathbf{r}^{\prime \prime }\right) ,  \label{ip1}
\end{eqnarray}%
it follows that if $\rho _{\lambda }\left( \mathbf{r}\right) =\lambda
^{3}\rho \left( \lambda \mathbf{r}\right) ,$ the uniformly scaled density,%
\begin{equation}
\frac{d}{d\lambda }\left. \frac{\delta F^{^{\gamma }}\left[ \rho _{\lambda }%
\right] }{\delta \rho _{\lambda }\left( \mathbf{r}\right) }\right\vert
_{\rho _{\lambda },\lambda =1}=\int d^{3}r^{\prime }\left( 3\rho \left(
\mathbf{r}^{\prime }\right) +\mathbf{r}^{\prime }\mathbf{.\nabla }\rho
\left( \mathbf{r}^{\prime }\right) \right) \frac{\delta ^{2}F^{^{\gamma }}%
\left[ \rho \right] }{\delta \rho \left( \mathbf{r}^{\prime }\right) \delta
\rho \left( \mathbf{r}\right) }.  \label{ip2}
\end{equation}%
Now consider the Schr\"{o}dinger equation%
\begin{eqnarray}
&&\left[ -\frac{1}{2}\sum_{i=1}^{N}\nabla _{i}^{2}+\frac{\gamma }{\lambda }%
\sum_{i<j}^{N}\frac{1}{\left\vert \mathbf{r}_{i}-\mathbf{r}_{j}\right\vert }%
+\sum_{i=1}^{N}v^{\frac{^{\gamma }}{\lambda }}\left( \left[ \rho \right] ;%
\mathbf{r}_{i}\right) \right] \Psi ^{\frac{^{\gamma }}{\lambda }}\left(
\left\{ \mathbf{r}_{i}\right\} \right)  \notag \\
&=&E\left( v^{\frac{^{\gamma }}{\lambda }}\left[ \rho \right] \right) \Psi ^{%
\frac{^{\gamma }}{\lambda }}\left( \left\{ \mathbf{r}_{i}\right\} \right)
\label{s1}
\end{eqnarray}%
from which it follows that%
\begin{eqnarray}
&&\left[ -\frac{1}{2}\sum_{i=1}^{N}\nabla _{i}^{2}+\gamma \sum_{i<j}^{N}%
\frac{1}{\left\vert \mathbf{r}_{i}-\mathbf{r}_{j}\right\vert }%
+\sum_{i=1}^{N}\lambda ^{2}v^{\frac{^{\gamma }}{\lambda }}\left( \left[ \rho %
\right] ;\mathbf{r}_{i}\right) \right] \Psi ^{\frac{^{\gamma }}{\lambda }%
}\left( \left\{ \lambda \mathbf{r}_{i}\right\} \right)  \notag \\
&=&\lambda ^{2}E\left( v^{\frac{^{\gamma }}{\lambda }}\left[ \rho \right]
\right) \Psi ^{\frac{^{\gamma }}{\lambda }}\left( \left\{ \lambda \mathbf{r}%
_{i}\right\} \right) .  \label{s2}
\end{eqnarray}%
The Levy constrained minimization approach \cite{Levy:79} implies that $\Psi
^{\frac{^{\gamma }}{\lambda }}\left( \left\{ \lambda \mathbf{r}_{i}\right\}
\right) $ yields $\lambda ^{3}\rho \left( \lambda \mathbf{r}\right) $ and
minimizes $\left\langle \Psi \left\vert \hat{T}+\gamma \hat{V}%
_{ee}\right\vert \Psi \right\rangle _{\Psi \rightarrow \lambda ^{3}\rho
\left( \lambda \mathbf{r}\right) }.$ Therefore \cite%
{ParrYang:bk89,DreizlerGross:bk90,JonesGunnarsson:89,LevyPerdew:85}%
\begin{equation}
\frac{\delta F^{^{\gamma }}\left[ \rho _{\lambda }\right] }{\delta \rho
_{\lambda }\left( \mathbf{r}\right) }+\lambda ^{2}v^{\frac{^{\gamma }}{%
\lambda }}\left( \left[ \rho \right] ;\lambda \mathbf{r}\right) =\mu
_{\lambda }^{\gamma }.  \label{ip3}
\end{equation}%
where, for an $N$-electron system
\begin{equation}
\mu _{N,\lambda }^{\gamma }=E_{N}^{\gamma }\left( \lambda ^{2}v^{\frac{%
^{\gamma }}{\lambda }}\left[ \rho _{N}\right] \right) -E_{N-1}^{\gamma
}\left( \lambda ^{2}v^{\frac{^{\gamma }}{\lambda }}\left[ \rho _{N}\right]
\right)  \label{ip3a}
\end{equation}%
with $E_{N}^{\gamma }\left( \lambda ^{2}v^{\frac{^{\gamma }}{\lambda }}\left[
\rho _{N}\right] \right) $ and $E_{N-1}^{\gamma }\left( \lambda ^{2}v^{\frac{%
^{\gamma }}{\lambda }}\left[ \rho _{N}\right] \right) $ the groundstate
energies of the $N$ and $\left( N-1\right) $ electron systems with the same
external potential $\lambda ^{2}v^{\frac{^{\gamma }}{\lambda }}\left[ \rho
_{N}\right] .$ From Eq.(\ref{s2}) it follows that
\begin{eqnarray}
\mu _{N,\lambda }^{\gamma } &=&E_{N}^{\gamma }\left( \lambda ^{2}v^{\frac{%
^{\gamma }}{\lambda }}\left[ \rho _{N}\right] \right) -E_{N-1}^{\gamma
}\left( \lambda ^{2}v^{\frac{^{\gamma }}{\lambda }}\left[ \rho _{N}\right]
\right)  \notag \\
&=&\lambda ^{2}\left( E_{N}^{\gamma }\left( v^{\frac{^{\gamma }}{\lambda }}%
\left[ \rho _{N}\right] \right) -E_{N-1}^{\gamma }\left( v^{\frac{^{\gamma }%
}{\lambda }}\left[ \rho _{N}\right] \right) \right)  \notag \\
&=&\lambda ^{2}\mu _{N}  \label{ip3b}
\end{eqnarray}%
where the last step follows from Eq. (\ref{e2}). The chemical potential $\mu
_{N}$ is independent of $\gamma $\cite{LevyGorling:96,LevyGorlingb:96}$.$
\bigskip Combining (\ref{e1}), (\ref{ip2}), (\ref{ip3}) and (\ref{ip3b})
yields (suppressing the subscript $N$ for convenience)%
\begin{eqnarray}
&&\int d^{3}r^{\prime }\left( 3\rho \left( \mathbf{r}^{\prime \prime
}\right) +\mathbf{r}^{\prime \prime }\mathbf{.\nabla }\rho \left( \mathbf{r}%
^{\prime \prime }\right) \right) \frac{\delta ^{2}F^{^{\gamma }}\left[ \rho %
\right] }{\delta \rho \left( \mathbf{r}^{\prime \prime }\right) \delta \rho
\left( \mathbf{r}\right) }  \notag \\
&=&2\mu -2v^{\gamma }\left( \left[ \rho \right] ;\mathbf{r}\right) -\mathbf{%
r.\nabla }v^{\gamma }\left( \left[ \rho \right] ;\mathbf{r}\right) -\left(
\gamma v_{x}\left( \left[ \rho \right] ;\mathbf{r}\right) +\gamma u\left( %
\left[ \rho \right] ;\mathbf{r}\right) +\gamma \frac{d}{d\gamma }v^{\gamma
}\left( \left[ \rho \right] ;\mathbf{r}\right) \right) .  \label{ip4}
\end{eqnarray}%
Upon substitution of (\ref{ip4}) into (\ref{v6}),

\bigskip
\begin{eqnarray}
&&\frac{\delta T^{\gamma }\left[ \rho \right] }{\delta \rho \left( \mathbf{r}%
\right) }+\frac{\delta F^{\gamma }\left[ \rho \right] }{\delta \rho \left(
\mathbf{r}\right) }+2v^{\gamma }\left( \left[ \rho \right] ;\mathbf{r}%
\right) +\left( \gamma v_{x}\left( \left[ \rho \right] ;\mathbf{r}\right)
+\gamma u_{x}\left( \left[ \rho \right] ;\mathbf{r}\right) +\gamma \frac{d}{%
d\gamma }v^{\gamma }\left( \left[ \rho \right] ;\mathbf{r}\right) \right)
\notag \\
&=&\int d^{3}r^{\prime }\frac{\delta T^{\gamma }\left[ \rho \right] }{\delta
\rho \left( \mathbf{r}^{\prime }\right) }f^{\gamma }\left( \mathbf{r}%
^{\prime }\right) +\int d^{3}r^{\prime }\frac{\delta F^{\gamma }\left[ \rho %
\right] }{\delta \rho \left( \mathbf{r}^{\prime }\right) }f^{\gamma }\left(
\mathbf{r}^{\prime }\right) -\int d^{3}r^{\prime \prime }\int d^{3}r^{\prime
}\mathbf{r}^{\prime }\mathbf{.\nabla }^{\prime }v^{\gamma }\left( \mathbf{r}%
^{\prime }\right) f^{\gamma }\left( \mathbf{r}^{\prime }\right) +2\mu
\label{v7}
\end{eqnarray}%
With the aid of (\ref{ip3}), (\ref{v3}) and (\ref{e1})
\begin{eqnarray}
&&\frac{\delta T_{c}^{\gamma }\left[ \rho \right] }{\delta \rho \left(
\mathbf{r}\right) }+\gamma \frac{d}{d\gamma }v_{c}^{\gamma }\left( \left[
\rho \right] ;\mathbf{r}\right) -v_{c}^{\gamma }\left( \left[ \rho \right] ;%
\mathbf{r}\right) -2\mu  \notag \\
&=&-\int d^{3}r^{\prime }\frac{\delta V_{ee}^{\gamma }\left[ \rho \right] }{%
\delta \rho \left( \mathbf{r}^{\prime }\right) }f^{\gamma }\left( \mathbf{r}%
^{\prime }\right) -2\int d^{3}r^{\prime }v^{\gamma }\left( \left[ \rho %
\right] ;\mathbf{r}^{\prime }\right) f^{\gamma }\left( \mathbf{r}^{\prime
}\right) -\int d^{3}r^{\prime }\mathbf{r}^{\prime }\mathbf{.\nabla }^{\prime
}v^{\gamma }\left( \left[ \rho \right] ;\mathbf{r}^{\prime }\right)
f^{\gamma }\left( \mathbf{r}^{\prime }\right)  \label{v8}
\end{eqnarray}%
where
\begin{equation}
T_{c}^{\gamma }\left[ \rho \right] =T^{\gamma }\left[ \rho \right] -T^{0}%
\left[ \rho \right] .  \label{v9}
\end{equation}%
From the definition of $E_{c}^{\gamma }\left[ \rho \right] $ and $%
T_{c}^{\gamma }\left[ \rho \right] $\cite%
{ParrYang:bk89,DreizlerGross:bk90,JonesGunnarsson:89,LevyPerdew:85}, Eqs. (%
\ref{ec1}) and (\ref{ec2}) it follows that%
\begin{equation*}
\frac{\delta T_{c}^{\gamma }\left[ \rho \right] }{\delta \rho \left( \mathbf{%
r}\right) }+\gamma \frac{d}{d\gamma }v_{c}^{\gamma }\left( \left[ \rho %
\right] ;\mathbf{r}\right) -v_{c}^{\gamma }\left( \left[ \rho \right] ;%
\mathbf{r}\right) =0,
\end{equation*}%
and therefore

\begin{eqnarray}
&&2\mu -2\int d^{3}r^{\prime }v^{\gamma }\left( \left[ \rho \right] ;\mathbf{%
r}^{\prime }\right) f^{\gamma }\left( \mathbf{r}^{\prime }\right) -\int
d^{3}r^{\prime }\mathbf{r}^{\prime }\mathbf{.\nabla }^{\prime }v^{\gamma
}\left( \left[ \rho \right] ;\mathbf{r}^{\prime }\right) f^{\gamma }\left(
\mathbf{r}^{\prime }\right)  \notag \\
&=&\int d^{3}r^{\prime }\gamma \frac{\delta V_{ee}^{\gamma }\left[ \rho %
\right] }{\delta \rho \left( \mathbf{r}^{\prime }\right) }f^{\gamma }\left(
\mathbf{r}^{\prime }\right)  \label{v10}
\end{eqnarray}%
The Fukui function is simply the difference between two densities \cite%
{PPLB:82}:%
\begin{equation}
f_{N}^{\gamma }\left( \mathbf{r}^{\prime }\right) =\rho _{N}^{1}\left(
\mathbf{r}\right) -\rho _{N-1}^{\gamma }\left( \mathbf{r}\right) .
\label{f1}
\end{equation}%
Note that by construction only the $\left( N-1\right) $-electron density
depends on $\gamma $. The external potential for the $N$ and $\left(
N-1\right) $ systems are the same, and hence the virial theorem \cite%
{LevyPerdew:85} and Eq. (\ref{f1}) imply that,%
\begin{eqnarray}
&&\int d^{3}r^{\prime }\mathbf{r}^{\prime }\mathbf{.\nabla }^{\prime
}v^{\gamma }\left( \left[ \rho _{N}^{1}\right] ;\mathbf{r}^{\prime }\right)
f^{\gamma }\left( \mathbf{r}^{\prime }\right)  \notag \\
&=&2T^{\gamma }\left[ \rho _{N}^{1}\right] +\gamma V_{ee}^{\gamma }\left[
\rho _{N}^{1}\right] -2T^{\gamma }\left[ \rho _{N-1}^{\gamma }\right]
-\gamma V_{ee}^{\gamma }\left[ \rho _{N-1}^{\gamma }\right] .  \label{v11}
\end{eqnarray}%
Combining (\ref{v10}), (\ref{v11}) and (\ref{ip3}) leads to the main result
of this paper:%
\begin{eqnarray}
&&V_{ee}^{\gamma }\left[ \rho _{N}^{1}\right] -V_{ee}^{\gamma }\left[ \rho
_{N-1}^{\gamma }\right]  \notag \\
&=&\int d^{3}r\frac{\delta V_{ee}^{\gamma }\left[ \rho _{N}^{1}\right] }{%
\delta \rho _{N}^{1}\left( \mathbf{r}\right) }\left( \rho _{N}^{1}\left(
\mathbf{r}\right) -\rho _{N-1}^{\gamma }\left( \mathbf{r}\right) \right) .
\label{v12}
\end{eqnarray}

\section{Corollary: Recursion relations}

Recursion relations can be derived at coupling strength interaction $\gamma $%
, when the single particle external potential is kept fixed, i.e. for an $M$%
-electron system
\begin{equation}
\hat{v}_{\text{ext}}^{M}=\sum_{i=1}^{M}v_{\text{ext}}^{\gamma }\left( \left[
\rho _{N}^{1}\right] ;\mathbf{r}_{i}\right) .
\end{equation}%
and $\rho _{M}^{\gamma }\equiv \rho _{M}^{\gamma }\left( v_{\text{ext}%
}^{\gamma }\left[ \rho _{N}^{1}\right] \right) $ is the $M$-electron density
constructed from a groundstate of the $M$-electron Hamiltonian $\hat{H}%
_{M}^{\gamma }=\hat{T}^{M}+\gamma \hat{V}_{ee}^{M}+\hat{v}_{M,\text{ext}%
}^{\gamma }\left[ \rho _{N}\right] .$ The potential $\hat{v}_{N,\text{ext}%
}^{1}\left[ \rho _{N}\right] $ for a real system is the interaction
potential between electrons and nuclei. From Eq. (\ref{v12}),
\begin{eqnarray}
&&V_{ee}^{\gamma }\left[ \rho _{N}^{1}\right] -V_{ee}^{\gamma }\left[ \rho
_{N-1}^{\gamma }\right]  \notag \\
&=&\int d^{3}r\frac{\delta V_{ee}^{1}\left[ \rho _{N}\right] }{\delta \rho
_{N}^{\gamma }\left( \mathbf{r}\right) }\left( \rho _{N}^{1}\left( \mathbf{r}%
\right) -\rho _{N-1}^{\gamma }\left( \mathbf{r}\right) \right)  \label{c1}
\end{eqnarray}%
and%
\begin{eqnarray}
&&V_{ee}^{\gamma }\left[ \rho _{N-1}^{\gamma }\right] -V_{ee}^{\gamma }\left[
\rho _{N-2}^{\gamma }\right]  \notag \\
&=&\int d^{3}r\frac{\delta V_{ee}^{\gamma }\left[ \rho _{N-1}^{\gamma }%
\right] }{\delta \rho _{N-1}^{\gamma }\left( \mathbf{r}\right) }\left( \rho
_{N-1}^{\gamma }\left( \mathbf{r}\right) -\rho _{N-2}^{\gamma }\left(
\mathbf{r}\right) \right) .  \label{c2}
\end{eqnarray}%
Continuing this pattern leads to
\begin{eqnarray}
&&V_{ee}^{\gamma }\left[ \rho _{N}^{\gamma }\right] -V_{ee}^{\gamma }\left[
\rho _{N-M}^{\gamma }\right]  \notag \\
&=&\sum_{L=0}^{N-1}\int d^{3}r\frac{\delta V_{ee}^{\gamma }\left[ \rho
_{N-L}^{\gamma }\right] }{\delta \rho _{N-L}^{\gamma }\left( \mathbf{r}%
\right) }\left( \rho _{N-L}^{\gamma }\left( \mathbf{r}\right) -\rho
_{N-1-L}^{\gamma }\left( \mathbf{r}\right) \right) ,  \label{c3}
\end{eqnarray}

and since $V_{ee}^{\gamma }\left[ \rho _{1}^{\gamma }\right] =0,$ it follows
that,

\begin{equation}
V_{ee}^{\gamma }\left[ \rho _{N}^{\gamma }\right] =\sum_{L=0}^{N-2}\int
d^{3}r\frac{\delta V_{ee}^{\gamma }\left[ \rho _{N-L}\right] }{\delta \rho
_{N-L}^{\gamma }\left( \mathbf{r}\right) }\left( \rho _{N-L}^{\gamma }\left(
\mathbf{r}\right) -\rho _{N-1-L}^{\gamma }\left( \mathbf{r}\right) \right) .
\label{c4}
\end{equation}

\section{Discussion and summary}

For $\gamma =0,$ $V_{ee}^{0}\left[ \rho \right] =E_{x}\left[ \rho \right] +U%
\left[ \rho \right] $\cite{ParrYang:bk89,DreizlerGross:bk90}. From Eq. (\ref%
{v12}) then follows the equation%
\begin{equation*}
E_{x}\left[ \rho _{N}^{1}\right] +U\left[ \rho _{N}^{1}\right] -E_{x}\left[
\rho _{N-1}^{0}\right] -U\left[ \rho _{N-1}^{0}\right] =\int d^{3}r\frac{%
\delta \left( E_{x}\left[ \rho _{N}^{1}\right] +U\left[ \rho _{N}^{1}\right]
\right) }{\delta \rho _{N}^{1}\left( \mathbf{r}\right) }\left( \rho
_{N}^{1}\left( \mathbf{r}\right) -\rho _{N-1}^{0}\left( \mathbf{r}\right)
\right)
\end{equation*}%
or%
\begin{eqnarray}
&&E_{x}\left[ \rho _{N}^{1}\right] -E_{x}\left[ \rho _{N-1}^{0}\right]
\notag \\
&=&\int d^{3}r\frac{\delta \left( E_{x}\left[ \rho _{N}^{1}\right] \right) }{%
\delta \rho _{N}^{1}\left( \mathbf{r}\right) }\left( \rho _{N}^{1}\left(
\mathbf{r}\right) -\rho _{N-1}^{0}\left( \mathbf{r}\right) \right)   \notag
\\
&&+\frac{1}{2}\int d^{3}rd^{3}r^{\prime }\left( \rho _{N}^{1}\left( \mathbf{r%
}\right) -\rho _{N-1}^{0}\left( \mathbf{r}\right) \right) \frac{1}{%
\left\vert \mathbf{r-r}^{\prime }\right\vert }\left( \rho _{N}^{1}\left(
\mathbf{r}^{\prime }\right) -\rho _{N-1}^{0}\left( \mathbf{r}^{\prime
}\right) \right) .
\end{eqnarray}%
This expression has already been derived by Levy and G\"{o}rling\cite%
{LevyGorlingb:95,LevyGorlingb:95}. The relation derived here, Eq. (\ref{v12}%
), is generalization of their work, valid at all coupling strengths. Note
that in the derivation of Eq. (\ref{vee}) use is made of Eq. (\ref{h1}), (%
\ref{s1}) and (\ref{s2}). This implies that derivation is correct for
v-representable densities \cite{ParrYang:bk89,DreizlerGross:bk90} and the
question whether it is correct for general densities remains.

In summary, a relationship between the Coulomb interaction functionals of a
many electron system at charge densities that differ by one electron,
evaluated at the same external potential, was derived. The derivation was
done for integer numbers of particles, but can be extended to fractional
particle numbers. As a corollary, it was shown that the Coulomb interaction
functional can be expressed as a sum over integrals of functional
derivatives of the Coulomb interaction functionals and charge densities for
all densities that differ form the total density by an integer. These
relations place stringent constraints on the energy functionals that appear
in density functional theory and it will be difficult for approximate
functionals to satisfy these equations.



\end{document}